\documentclass[oribibl]{llncs} 
\usepackage{graphicx} 
\usepackage{subfig}
\pdfoutput=1
\usepackage{amsmath}
\usepackage{oz}
\usepackage{color}
\usepackage{stmaryrd}
\usepackage{graphicx}
\usepackage[utf8x]{inputenc}

\newcommand{\sem}[1]{\llbracket #1 \rrbracket}

\newcommand{\mc}[1]{\mathcal{#1}}
\newcommand{\linefill}{\cleaders\hbox{$\smash{\mkern-2mu\mathord-\mkern-2mu}$}\hfill\vphantom{\lower1pt\hbox{$\rightarrow$}}}  
\newcommand{\Linefill}{\cleaders\hbox{$\smash{\mkern-2mu\mathord=\mkern-2mu}$}\hfill\vphantom{\hbox{$\Rightarrow$}}}  
\newcommand{\transi}[2]{\mathrel{\lower1pt\hbox{$\mathrel-_{\vphantom{#2}}\mkern-8mu\stackrel{#1}{\linefill_{\vphantom{#2}}}\mkern-11mu\rightarrow_{#2}$}}}
\newcommand{\trans}[1]{\transi{#1}{{}}}
\newcommand{\transo}{\mathord{\trans{~}}}
\newcommand{\ntransi}[2]{\mathrel{\lower1pt\hbox{$\mathrel-_{\vphantom{#2}}\mkern-8mu\stackrel{#1}{\linefill_{\vphantom{#2}}}\mkern-8mu\nrightarrow_{#2}$}}}

\newcommand{\Transi}[2]{\mathrel{\lower1pt\hbox{$\mathrel=_{\vphantom{#2}}\mkern-8mu\stackrel{#1}{\Linefill_{\vphantom{#2}}}\mkern-11mu\Rightarrow_{#2}$}}}

\newcommand{\cupdot}{\stackrel{.}{\cup}}

\newcommand{\false}{\mathbf{0}}
\newcommand{\maybe}{\mbox{\textbf{\textonehalf}}}
\newcommand{\true}{\mathbf{1}}

\DeclareMathOperator{\id}{id}

\newcommand{\br}[1]{\left(#1\right)}
\newcommand{\cbr}[1]{\left\{#1\right\}}

\newcommand{\abr}[1]{\left\lseq #1\right\rseq}
\newcommand{\abs}[1]{\left\lvert #1\right\rvert}
\newcommand{\ra}{\rightarrow}
\newcommand{\raa}{\longrightarrow}

\newcommand{\LRa}{\Leftrightarrow}
\newcommand{\lio}{\sqsubseteq}

\newcommand{\aux}{\mathcal{I}}
\newcommand{\core}{\mathcal{C}}
\newcommand{\unary}{\core^1}
\newcommand{\binary}{\core^2}

\graphicspath{{figures/}}

\begin{document}

\title{Towards a Shape Analysis for\\Graph Transformation Systems}

\author{Dominik Steenken, Heike Wehrheim, Daniel Wonisch}

 \institute{Universit\"at Paderborn \\ Institut f\"ur Informatik \\ 33098 Paderborn, Germany \\
            {\small \{dominik,wehrheim,dwonisch\}@mail.uni-paderborn.de}}

\maketitle
 
\begin{abstract}
   Graphs and graph transformation systems are a frequently used
modelling technique for a wide range of different domains, covering
areas as diverse as 
refactorings, network topologies or reconfigurable software. 
Being a formal method, graph transformation systems lend themselves
to a formal analysis. This has
inspired the development of various verification methods,
in particular also model checking tools. 

\smallskip
In this paper, we present a verification technique for {\em infinite-state}
graph transformation systems. The technique employs
the abstraction principle used in shape analysis of programs,
summarising possibly infinitely many nodes thus giving {\em shape 
graphs}. The technique has been implemented using the $3$-valued
logical foundations of standard shape analysis. We exemplify the
approach on an example from the railway domain. 

\end{abstract}
 
 \section{Introduction}
 
 Graph transformation systems (GTSs, \cite{CorradiniMREHL97}) have - in particular due to their visual appeal - become a widely used technique for system modelling. They are employed in numerous different areas, ranging from the specification of visual contracts for software to dynamically evolving systems. They serve as a formally precise description of the behaviour of complex systems. Often, such systems are operating in safety critical domains (e.g.\ railway, automotive) and their dependability is of vital interest. Hence, a number of approaches for the analysis of graph transformation systems have been developed \cite{Taentzer03}, in particular also model checking techniques \cite{RensinkSV04,SchmidtV03,Rensink03,BaldanCK08,BauerBKR08,SWJ2008}. Model checking allows to fully automatically show properties of system models, for instance for properties specified in temporal logic. Model checking proceeds by exploring the whole state space of a model, i.e.\ in case of graph transformation systems by generating the set of graphs which are reachable from a given start graph by means of rule application. While existing tools have proven to be able to tackle also large state spaces, standard model checking techniques fail when the state space becomes infinite.

There are, in general, two approaches to dealing with very large or even infinite state spaces. The first approach is to devise a clever way of selecting a \emph{finite subset} of states which is sufficient for proving the desired properties, effectively constructing an \emph{under-approximation} of the system. This concept is explored e.g. in \emph{bounded model checking} \cite{biere2003bounded}. The second approach is to construct an \emph{abstraction}, i.e. a finite representation of a \emph{superset} of the state space. This \emph{over-approximation} of the system is then used to show certain properties of the original system.

In this paper, we propose an new approach towards a verification technique for infinite state graph transformation systems using over-approximation. The technique follows the idea of {\em shape analysis} algorithms for programs \cite{SagivRW02} which are used to compute properties of a program's heap structures. Shape analyses compute abstractions of heap states by collapsing certain sets of identical nodes into so-called summary nodes. Thereby, an infinite number of heap states can be finitely represented. Such {\em shapes} can be used to derive structural properties about the heap. 

This principle of summarisation in GTSs has already been presented in numerous other works, for example Rensink et al.\  \cite{Rensink04,RensinkD06} or Bauer et.al.\ \cite{BauerBKR08} which introduce so-called abstract graph transformations. Our approach is set apart from these by strict adherence to the formalism presented in \cite{SagivRW02}, which immensely simplifies implementation and gives us a level of parametrization that other approaches lack. A more thorough discussion of advantages of our approach over related work will be presented in section \ref{sec:conclusion}.

Here, we present a shape analysis for GTSs which is directly based on the $3$-valued logical foundations of standard shape analysis. Given this logical basis for shape graphs, we define rule application on shape graphs via a {\em constructive} definition of materialisation and summarisation. The technique can thus be directly implemented as defined, even re-using parts of the logical machinery of TVLA \cite{BogudlovLRS07}, the most prominent shape analysis tool. In order to illustrate our technique, we exemplify it on a simple GTS model from the railway domain.

The paper is structured as follows. The next section will give the basic definitions for our approach. Section~\ref{sec:method} introduces materialisation and summarisation on graphs and thereby defines the application of rules on shape graphs. Section~\ref{sec:result} shows the correctness of our approach, i.e.\ shows that by rule application on shape graphs an overapproximation of the set of reachable graphs is computed. The next section then reports on the implementation. Finally, Sect.~\ref{sec:conclusion} concludes, further discusses related work and gives some directions for future research.

 \section{Background}
 
 % Contains subsections ``Graph Transformation Systems'', ``3-valued Logic'' and ``Shape Graphs''
 This section introduces the basic definitions that are required to
formulate our main results. To illustrate the definitions in
this section, we use the following example from the rail domain.  
A rail network is given by a set of stations (S) and a set of rail
sections, called tracks (T), connected by a  relation called
``next''. Vehicles, called 
``railcabs'' (RC), possibly with passengers (P) travel on the
tracks. The example is a  
simplified version of a case study 
coming from the project ``Neue Bahntechnik
Paderborn''\footnote{http://nbp-www.upb.de}. 
\begin{figure}
 \centering
 \includegraphics[width=5.5cm]{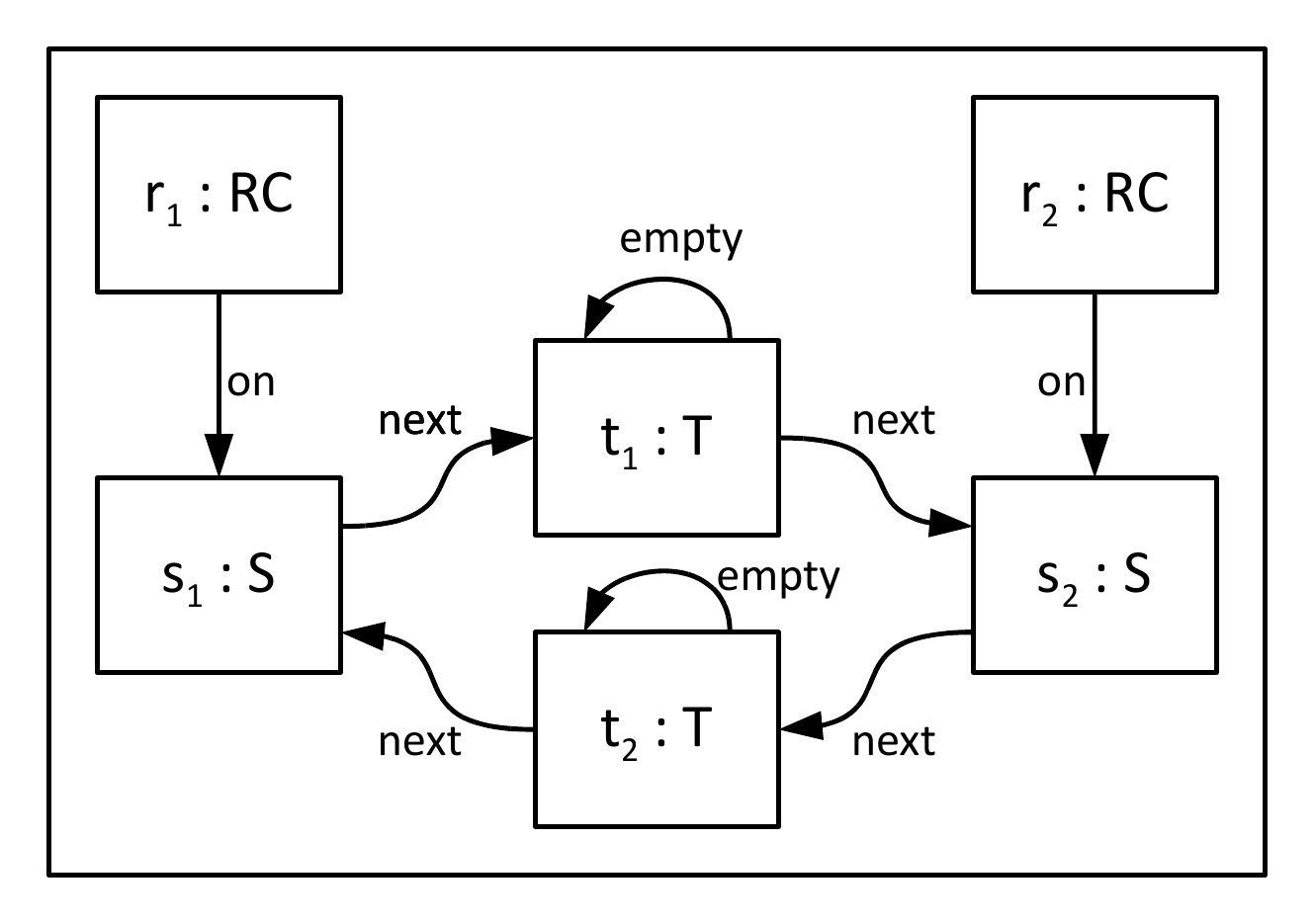}
 \caption{A simple rail network}
 \label{fig:simplerail}
 \vspace{-1.5em}
\end{figure}
Figure
\ref{fig:simplerail} shows a graph depicting one configuration of
such a rail network. Configurations can change in a number of
ways, for instance by passengers entering railcabs and railcabs moving
on tracks according to predefined protocols. The overall goal is to
show certain safety properties (e.g.\ collision avoidance) for
arbitrary networks. We first of all start by defining some basic notions on graphs.

\begin{definition}
 A {\em graph} $G$ is a pair $(N,E)$, where $N$ is a set of {\em
 nodes}  and $E \subseteq N \times {\cal L} \times N$ is a set of
 {\em labelled edges} for some label set $\cal L$.
 For any graph $G$, $N_G$ and $E_G$ denote its node and edge sets, respectively.
\end{definition}

\noindent This definition restricts the class of graphs we are considering to
those in which no more than a single same-labelled edge may exist between any two
nodes. The generic concept of a morphism extends to these graphs in a
natural way. 

\begin{definition}
 For graphs $G$ and $H$, a morphism $f: G\ra H$ is a function
 $f: N_G \ra N_H$ extended to edges by $f\br{n,l,n'} = \br{f(n),l,f\br{n'}}$
 such that $f\br{E_G} \subseteq E_H$.
\end{definition}

\noindent Figure~\ref{fig:simplerail} shows a graph representing one
very simple rail network consisting of two stations which are
connected by two tracks. % such that the resulting structure forms a
%cycle. 
Note that we include a simple notion of typing in the
graph. The type of a node is represented by a loop labelled with the
name of the type. Such type loops are not displayed as edges but
rather as part of the node name. Thus, instead of displaying a
self-edge of $r_1$ labelled ``RC'', we label the node $r_1:RC$. 

In order to model the dynamic behaviour of a system represented by a
graph, we need to transform graphs into other graphs. For this, graph
production rules can be used. In this paper, we take an operational,
not categorical, view on graph transformation. As a consequence, we
favour a simple approach to graph production rules, as the following
definitions show. 

\begin{definition} A {\em graph production rule} $P = \abr{L,R}$
  consists of two graphs $L$ and $R$ called the left hand side and the
  right hand side, respectively.
\end{definition}

\begin{figure}[t]
 \centering
 \includegraphics[width=7.5cm]{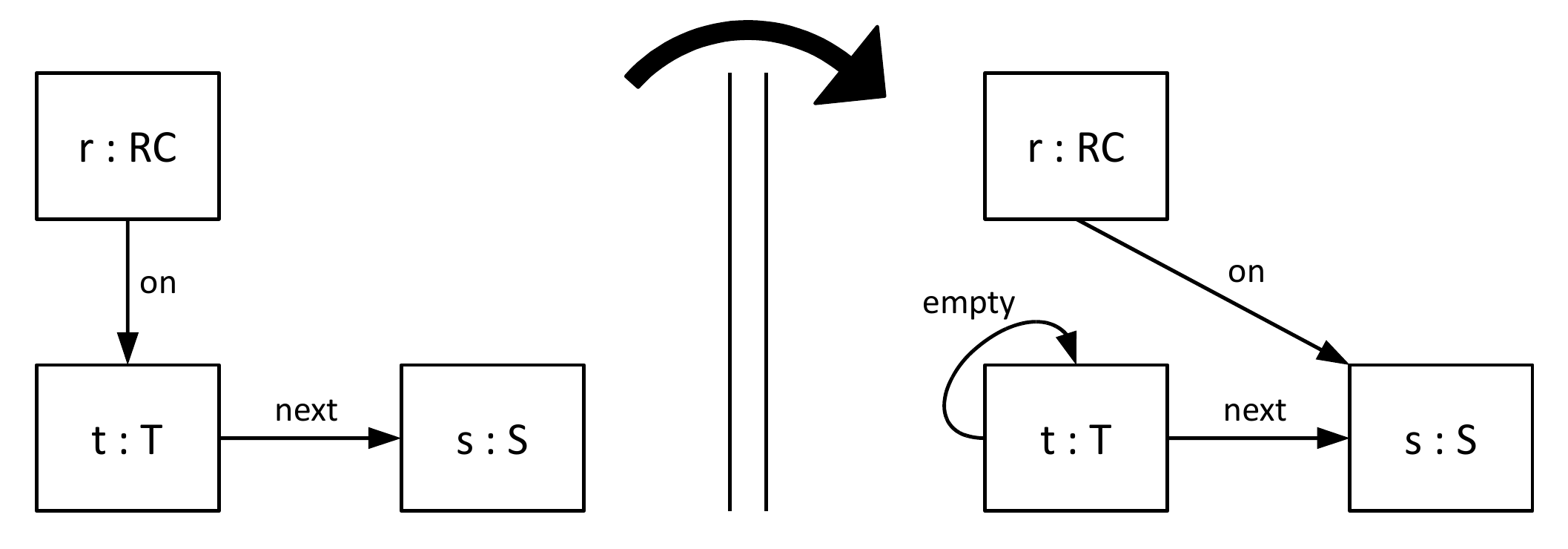}
 \caption{Rule $EnterStation$}
 \label{fig:moverule}
 \vspace{-1.5em}
\end{figure}

\noindent Figure~\ref{fig:moverule} shows an example of a rule which
describes a railcab entering a station. In addition, we
have rules for leaving a station, for movement of single
as well as {\em convoys} of railcabs and for forming convoys
(all elided due to space restrictions). 
In the rules, we
use node names instead of injective morphisms  to identify nodes
appearing in the left as well as right hand side. Aside from this
technicality we use the standard SPO approach to rule definition and
application \cite{Lowe93}. In order to make node creation and deletion
explicit, we use the following sets: 
\begin{align*}
  N^{-} = N_L \setminus N_R,\ & E^{-} = E_L \setminus E_R
  & \mbox{(deleted nodes and edges)}\\ 
  N^{+} = N_R \setminus N_L,\ & E^{+} = E_R \setminus E_L
  & \mbox{(created nodes and edges)} 
\end{align*}

\noindent These sets are used to define the effect of an application
of a production rule on a graph $G$. 

\begin{definition}
Let $P$ be a production rule, $G$ a graph.  The rule
$P=\abr{L,R}$  {\em  can be applied on G} if we can find an injective
morphism $m : L 
\ra G$ (called a {\em matching}).\\
If $m$ is a matching, then the application of $P$ onto $G$ with
matching $m$ is the graph
\begin{align*}
  H   & = \br{N_H,E_H}\ \mbox{with}\\
  N_H & = \br{N_G \setminus m\br{N^{-}}} \cupdot N^{+}\\
  E_H & = \br{\br{E_G \setminus m\br{E^{-}}} \cup \widehat{m}\br{E^{+}}} \cap \br{N_H \times \mc L \times N_H}
\end{align*}

where $\widehat{m} = m \cup \id_{N^{+}}$.
\end{definition} 

\noindent For this production application we write $G \trans{P, m}
H$. Similarly, $G \trans P H$ holds if there is some $m$ such that $G 
\trans {P, m} H$ and $G \raa H$ if there is furthermore a production
rule $P$ such that $G \trans P H$. We let $\transo^*$ denote the
transitive and reflexive closure of $\transo$. With these definitions
at hand, we can define the set of reachable graphs of a graph
transformation system (or {\em graph grammar}, as it includes a start graph). 
 
\begin{definition}
 A {\em graph transformation system} $GT = (G_0, \br{P_i}_{i \in I})$ consists of a
 start graph $G_0$ and a set of production rules $P_i, i\in I$. 
 The set of {\em reachable graphs} of a graph transformation system
 $GT$ is 
 \[ reach(GT) = \{ G \mid G_0 \trans {}^* G \} \]
\end{definition} 

\noindent 
In this paper we are interested in proving properties of the
set of all reachable graphs. A property can for instance be the
absence of {\em forbidden patterns}, i.e.\ substructures, in a
graph (or the presence of desired patterns).  
\begin{sidebyside}
Such a forbidden pattern can be defined by a
production rule of the form $P=\abr{F,F}$ (with left and right hand
equal). The pattern is {\em present} in a graph $G$ ($G \models F$) if the rule
matches. A forbidden pattern for our example is given on the right
hand side.

It specifies a collision of two railcabs
(two railcabs on 
one track). 
\nextside

\vspace*{-.8cm} \begin{center}
 	\includegraphics[scale=0.45]{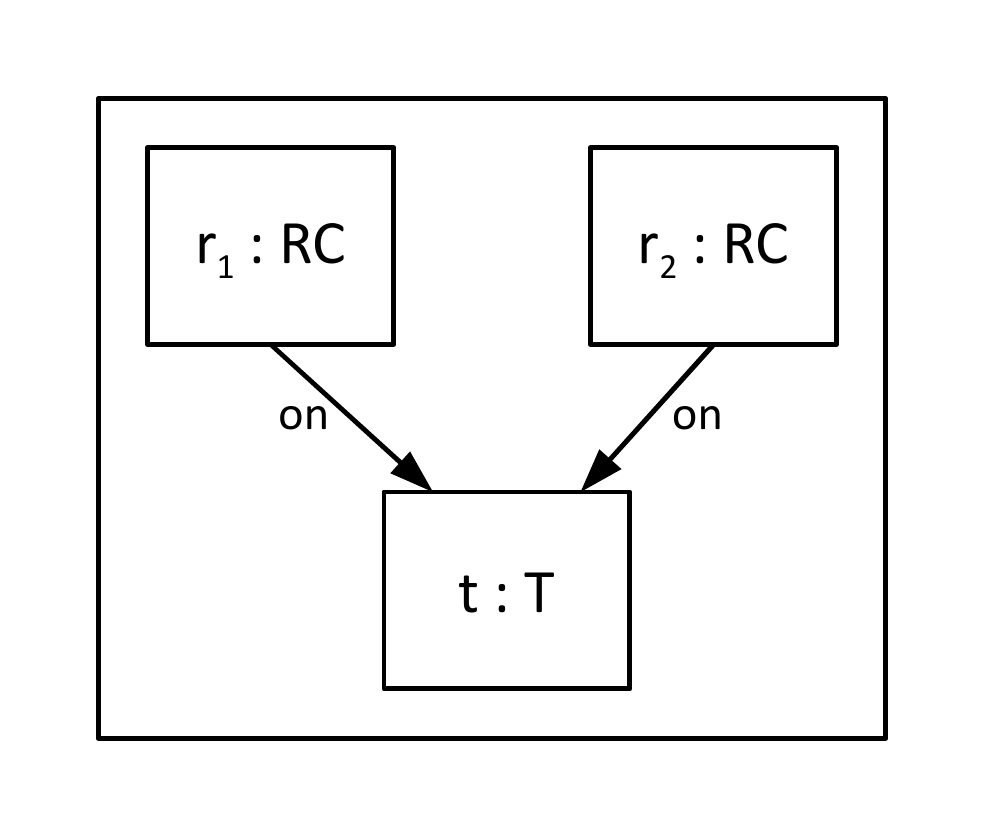}
 \end{center}
\end{sidebyside}

\vspace*{-.8cm}
\noindent The set of reachable graphs can in general be infinite
(e.g.\ for our example, if we introduce a rule which allows new passengers to
be created). The objective of this paper is to construct an
{\em abstraction} (and overapproximation) of this set of reachable
graphs which is finite but on 
which we can still show properties. 

Before doing so, we need to look a bit closer into the basic
technology behind shape analysis. Shape analysis algorithms
operate on logical a structure using first 
order logic to formulate properties. In the following, we closely
follow \cite{SagivRW02} in our notations. Note, however, that we explicitly exclude the notion of transitive closure from \cite{SagivRW02}, since transitivity would violate the important locality property of rule applications. The word {\em formula} always
refers to a first order formula over a set of predicate symbols $\mc
P$ and variables $\mc V$. Variables are assigned values from some
domain (or universe) $U$, and $k$-ary predicates $\mc P_k$ are
interpreted by truth-valued functions, i.e.\ we have an interpretation function 
$\iota: {\mc P_k} \rightarrow (U^k \rightarrow {\mc T})$ (
$\mc T$ a set of truth values). We let $\mc F\br{\varphi}$
denote the set of free variables of a formula $\varphi$. 
Domain,
predicates and interpretation function together make up a logical
structure $S=\abr{U,\mc P,\iota}$, sometimes also abbreviated by
$\abr{U,\iota}$. For a formula $\varphi$, $\sem{\varphi}_{m}^S$
denotes the value of $\varphi$ in the structure $S$ under an
assignment $m$. 

\begin{figure}
  \begin{center}
    \includegraphics[scale=0.55]{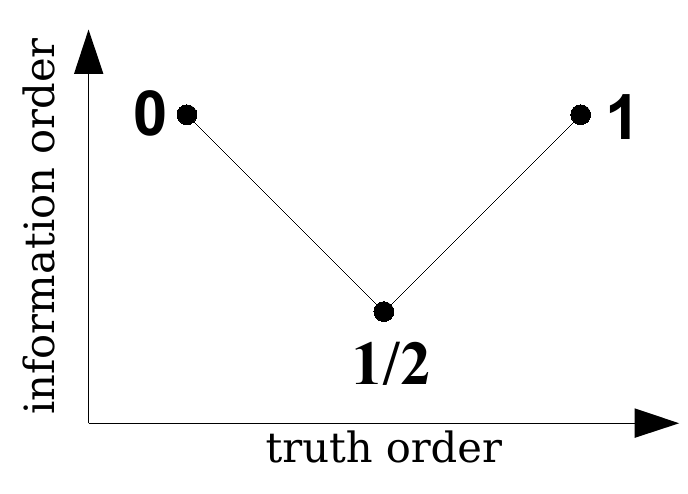}
  \end{center}  
  \vspace{-1em}
  \caption{Kleene logic with logical and information order}
  \label{fig:kleene}
  \vspace{-1.5em}
\end{figure}

A logical structure is called \emph{$n$-valued} if for the target set
$\mc T$ of the predicates $\abs{\mc T}=n$ holds. Two sets of truth
values will play a role here: the ordinary boolean values ($\mc T =
\{\false,\true\}$) and the three-valued set of Kleene logic ($\mc T =
\{\false,\true,\maybe\}$, values called $false$, $true$ and $maybe$). 
On the truth values of Kleene logic we have two different orderings (see
Fig.~\ref{fig:kleene}), one reflecting the amount of {\em information}
($\lio$) 
present in a logical value, the other the {\em logical truth} ($\leq$). That is, for $l_1,l_2\in\mc T=\cbr{\false,\maybe,\true}$:
\begin{align*}
  l_1\lio l_2 & \LRa \br{l_1=l_2}\vee\br{l_2=\maybe} \\
  l_1\leq l_2 & \LRa \br{l_1=l_2} \vee \br{l_1=\false} \vee \br{l_1=\maybe \wedge l_2=\true}
\end{align*}

\smallskip
\noindent 
Our final goal is to represent graphs as well as their abstractions by
logical structures, the former by 2-valued and the latter by
3-valued. To this end, we partition 
 the set $\mc P$ into the sets of so-called \emph{core predicates}
 $\core$ and \emph{instrumentation predicates} $\aux$. Later on,
 $\core$ will encode basic properties (like $next$-relations between
 nodes), while $\aux$ will be used to 
 increase the precision of the analysis with respect to a given
 property. The set $\core$ is further subdivided into 
 \emph{unary core predicates} $\unary$ (used e.g.\ for types) and \emph{binary
   core predicates} $\binary$. One specific predicate called {\em
   summarised} ($sm$) is used in the abstraction: a summarised node
 can represent lots of concrete nodes. In ordinary graphs no nodes are
 summarised. 

The encoding of graphs as logical structures then works as follows:
The set of nodes $N$ of a 
graph will be represented by the domain set $U$. The edge labels $\mc
L$ will give us the set of predicate symbols $\mc P$, and particular
edges are encoded by $\iota$. 
Table~\ref{tab:graph_encoding}
gives the logical structure of the rail network of Fig.~\ref{fig:simplerail}. 

\begin{table}[t]
 \begin{center}
    \begin{align*}
    U & =\cbr{r_1,r_2,s_1,s_2,t_1,t_2}\\
    \unary & = \cbr{RC, T, S, sm}, \binary = \cbr{on,next} 
   \end{align*}
   \begin{tabular}{| c ||c|c|c|c|c| c |c|c|c|c|c|c| c |c|c|c|c|c|c|c|}
    \hline
    $\iota\br{\unary}$ & $RC$ & $T$ & $S$ & $sm$ & $\iota\br{on}$ & $r_1$ & $r_2$ & $s_1$ & $s_2$ & $t_1$ & $t_2$ & $\iota\br{next}$ & $r_1$ & $r_2$ & $s_1$ & $s_2$ & $t_1$ & $t_2$ \\
    \hline
    \hline
    $r_1$ & $1$ & $0$ & $0$ & $0$ & & $0$ & $0$ & $1$ & $0$ & $0$ & $0$ & & $0$ & $0$ & $0$ & $0$ & $0$ & $0$\\
    \hline
    $r_2$ & $1$ & $0$ & $0$ & $0$ & & $0$ & $0$ & $0$ & $1$ & $0$ & $0$ & & $0$ & $0$ & $0$ & $0$ & $0$ & $0$\\
    \hline
    $s_1$ & $0$ & $0$ & $1$ & $0$ & & $0$ & $0$ & $0$ & $0$ & $0$ & $0$ & & $0$ & $0$ & $0$ & $0$ & $1$ & $0$\\
    \hline
    $s_2$ & $0$ & $0$ & $1$ & $0$ & & $0$ & $0$ & $0$ & $0$ & $0$ & $0$ & & $0$ & $0$ & $0$ & $0$ & $0$ & $1$\\
    \hline
    $t_1$ & $0$ & $1$ & $0$ & $0$ & & $0$ & $0$ & $0$ & $0$ & $0$ & $0$ & & $0$ & $0$ & $0$ & $1$ & $0$ & $0$\\
    \hline
    $t_2$ & $0$ & $1$ & $0$ & $0$ & & $0$ & $0$ & $0$ & $0$ & $0$ & $0$ & & $0$ & $0$ & $1$ & $0$ & $0$ & $0$\\
    \hline
   \end{tabular} 
 \end{center}
 \caption{Logical structure of the graph in Figure~\ref{fig:simplerail}}
 \label{tab:graph_encoding}
 \vspace{-1.5em}
\end{table}

\begin{definition}
 Let $G$ be a graph. The \emph{$2$-valued encoding} of $G$, denoted
 $ls\br{G}$, is a $2$-valued logical structure $S=\abr{U,\mc P,\iota}$
 with $U = N_G$, $\unary \cup \binary = \mc P = \mc L$ 
 and $\iota$ defined by:
 \begin{itemize}
  \item For $p\in\binary$: $\iota\br{p}\br{u_1,u_2} =
 1 \LRa \br{u_1,p,u_2}\in E_G$, 
  \item For $p\in\unary$: $\iota\br{p}\br{u} = 1 \LRa \br{u,p,u}\in
 E_G$, 
  \item For $sm$: $\iota\br{sm}\br{u} = 0$. 
 \end{itemize}

\end{definition}

\begin{figure}[t]
  \begin{center}
   \includegraphics[scale=0.5]{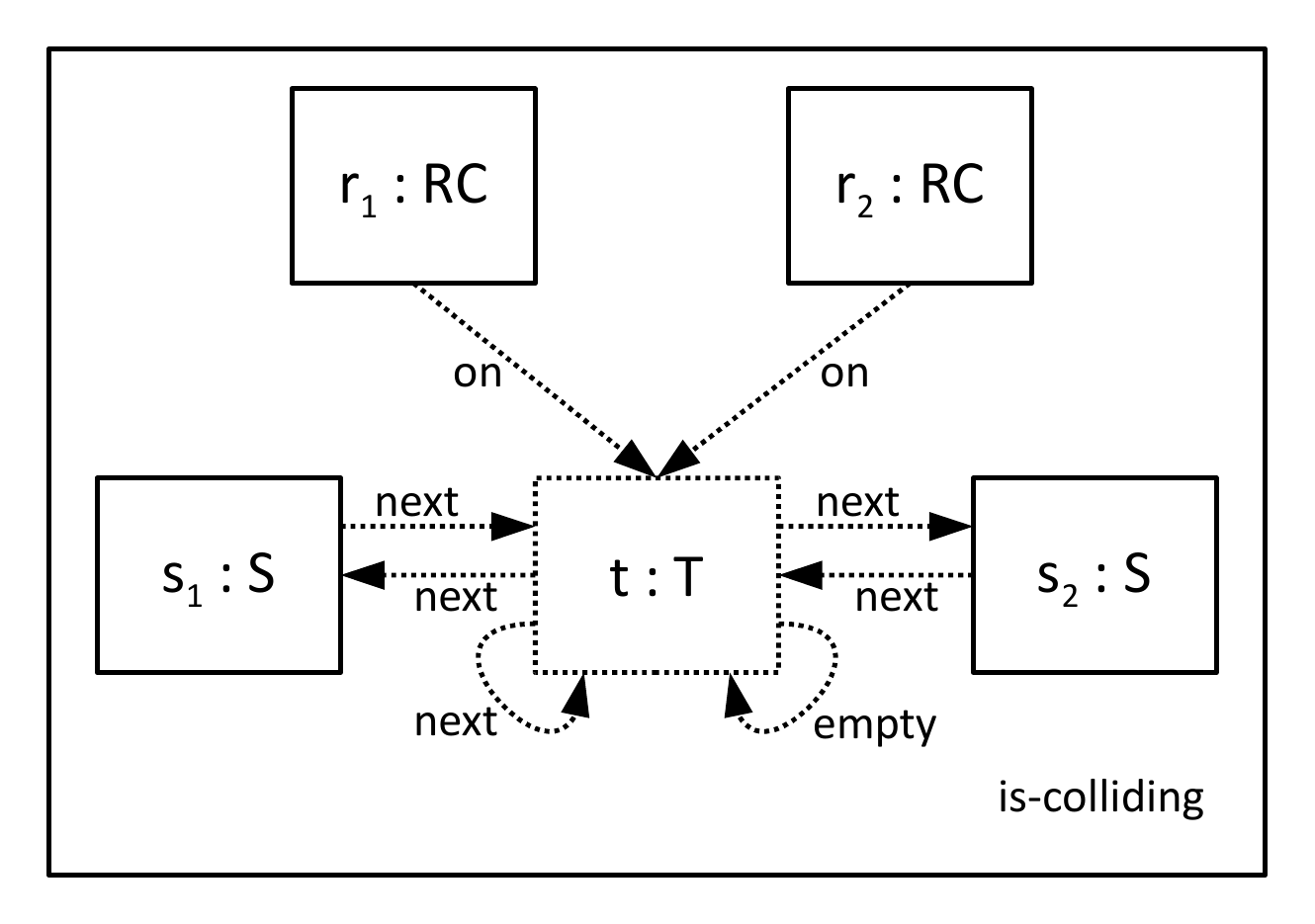}
   \end{center}
 \vspace{-2.0em}
 \caption{Start shape graph}
 \label{fig:shape_graph_encoding}
 \vspace{-1.5em}
\end{figure}

\noindent The basic idea of shape analysis is to represent infinitely
many different {\em but in shape similar} configurations or graphs by one {\em
  shape graph}. A shape graph thus cannot always give us precise information
about the number of nodes, nor can it give us precise information about
edges between nodes. The third truth value ``maybe''
($\maybe$) is used to represent  this fact in logical
structures. A node representing many concrete nodes is {\em
  summarised}, denoted by a dashed rectangle, an edge which is only
``maybe'' there (dashed line) is assigned the truth value
$\maybe$. Figure \ref{fig:shape_graph_encoding} shows a shape
  graph. Here, we for instance have $RC(r_1) = \true$ ($r_1$ is definitely
  of type railcab), $sm(t) = \maybe$ (there is possibly more than one 
  track) and
  $next(t,t) = \maybe$ (the tracks summarised in $t$ {\em maybe} 
  connected). The predicate $is\_colliding$ will be explained
  later. This is the start shape graph of our reachability 
  analysis. 
% and its
%3-valued logical structure. 

Shape graphs are abstractions of concrete graphs, concrete
graphs can be {\em embedded} into them.  Clearly , the predicate $sm$ plays
a crucial role in embeddings. Interpretations of this predicate are
restricted to values $\false$ and $\maybe$. If it is $\maybe$ for
an individual $u$, this means $u$ may or may not stand for a whole
set of nodes. If it is $\false$ for $u$, then $u$ is guaranteed to
represent a single individual. 

We thus obtain the notion of embedding
by the following definition. 

\begin{definition}
 Let $S=\abr{U,\mc P,\iota},S'=\abr{U',\mc P,\iota'}$ be two logical
 structures and $f:U\ra U'$ be a surjective function. We say that $f$
 \emph{embeds} $S$ in $S'$ ($S\lio S'$) iff $\forall k\forall p\in\mc{P}_k,\forall
 u_1,\ldots,u_k\in U$ 
 \begin{align}
   \iota\br{p}\br{u_1,\ldots,u_k}\lio\iota'\br{p}\br{f\br{u_1},\ldots,f\br{u_k}} \label{eq:embed_condition_1}\end{align}   
 and $\forall u'\in U'$ 
 \begin{align} \br{\abs{\cbr{u\mid
         f\br{u}=u'}}>1}\lio\iota'\br{sm}\br{u'} \label{eq:embed_condition_2}\end{align} 
 %We write $S\lio S'$ to denote that there is some surjective function
 %$f$ which embeds $S$ in $S'$. 
\end{definition}

\noindent Thus, intuitively, $S\lio S'$ means that $S'$ is in some way a
``generalisation'' of $S$. 
%Thus, by means of the embedding function
%$f$, $S'$ ``represents'' $S$. Since there may be infinitely many other
%logical structures which embed into $S'$, the term generalisation is
%appropriate. 

\section{Rule Application on Shape Graphs}
 \label{sec:method}
 The basic idea of shape analysis follows that of abstract
interpretation: instead of looking at concrete graphs and applying graph
transformation rules concretely, we look at shape graphs and apply
our rules to shapes instead. We thus inductively compute the {\em set
of reachable shape graphs} 
and on these check for forbidden patterns. If the forbidden pattern is
absent in this set, it should also not be present in any concretely
reachable graph. In this section we will explain how to apply rules on
shape graphs, the next section will look at the soundness of this
technique.

Rule application on shape graphs involves a number of distinct steps,
some of which are not present on concrete graphs. The basic difference
is that due to the ``maybe'' predicates in shapes, we usually do not find an exact counterpart, i.e. an injective matching, for
the left hand side.  The
following steps are necessary: 
\begin{description}
        \item[Match] To find out whether a rule $P$ 
        matches, we evaluate a {\em rule formula} $\varphi_P$ in the logical
        structure of the shape graph. If it evaluates to $\maybe$,
        the rule can potentially be applied. 
        \item[Focus] In order to actually apply a potentially
        applicable rule, we have to bring the left hand side of the
        rule into focus. We do so by {\em materialising} the left hand
        side in the shape graph. 
         \item[Coerce] Materialisation concretises parts of the shape
        graph. This concretisation has an influence on the rest of the
        shape (e.g., if a railcab is definitely on one track, it
        cannot at the same time be ``maybe'' on another
        track). Coercing removes ``maybe'' structures in the shape by inspecting
        definitely known predicates. 
         \item[Apply] After materialisation and coercion the
        rule can be applied, basically as on concrete graphs. 

\end{description} 

\noindent We next go through each of these steps. To
        define matching, we transform the left hand side of a rule into a
        formula. 

\begin{definition}
  Let $P=\abr{L,R}$ be a graph production rule. The \emph{production
  formula} $\varphi_P$ corresponding to $P$ is given by 
  \begin{align*}
   \varphi_P = & \underbrace{\bigwedge_{\substack{(n,l,n')\in
   E_L\\l\text{
   binary}}}l\br{n,n'}}_{\text{edges}}\wedge\underbrace{\bigwedge_{\substack{\br{n,l,n}\in
   E_L\\l\text{
   unary}}}l\br{n}}_{\text{loops}}\wedge\underbrace{\bigwedge_{\substack{n_1,n_2\in
   N_L\\n_1\neq
   n_2}}\neg\br{n_1=n_2}}_\text{injectivity}\wedge \underbrace{\bigwedge_{n\in
   N_L}\neg sm\br{n}}_\text{non-summarisation} 
  \end{align*}
  The $\neq$ here means the non-equality of the \emph{variable
 symbols}, while $=$ is a regular predicate\footnote{Given an
 assignment $m$, two variables $x_1$ and $x_2$ are considered
 equal if they are mapped onto the same node by $m$ and this is not a summary
 node.}. 
 \end{definition} 

\noindent %This formula is satisfiable in the 2-valued logical structure of a
 %concrete graph if and only if the rule is applicable on the
 %graph, and the assignment $m$ of variables to individuals then gives
 %us a matching.  
 When this formula evaluates to $\maybe$ ($\true$) for a 3-valued
 structure, we say that the rule is {\em potentially applicable}
 (applicable) in the associated shape graph. To actually apply it, we
 have to bring the rule into {\em focus}, i.e.\ make sure that we definitely
 find the left hand side of the rule in the shape. 
Intuitively, we would want something like this
 \begin{align*}
  focus_P\br{S}=\cbr{S'\mid S'\lio S\wedge \exists m: \sem{\varphi_P}_m^{S'} = \true}
 \end{align*}
 meaning all possible graphs which are embeddable in $S$ and to which
 the rule can be applied. Unfortunately, this set can be infinitely
 large, and in fact, this is exactly what our technique tries to
 avoid, namely having to construct {\em all} concrete graphs for a shape. 
 Instead, we only compute a set $mat_P\br{S}$ (the {\em
 materialisation} with respect to a rule) such that each
 element in $focus_P\br{S}$ can be embedded in at least one element
 from $mat_P\br{S}$, but still the rule is applicable in every shape
 in  $mat_P\br{S}$. 
 
 In order to construct the set $mat_P\br{S}$, let us now assume that
 we have a shape graph $G$, its corresponding logical structure
 $S=\abr{N_G,\mc L,\iota}$, a production rule $P=\abr{L,R}$,
 and a matching $m:L\ra N_G$ which gives rise to an assignment $\bar
 m:\mc F\br{\varphi_P} \ra N_G$ such that $\sem{\varphi_P}_{\bar m}^S\neq\false$. 
Let $N_G^{sum}$ be the set of summary nodes in
 $G$. %If no summary nodes are in the range of $m$, then the
% applicability of the rule using $m$ can be determined as
 %usual. Therefore, we now assume that $\Gamma\br{m}:=m(L)\cap
 %N_G^{sum}\neq\emptyset$. 
We have  to exactly
 find the left hand side of the rule in the materialisation. Thus,
 every node $u$ in $\Gamma(m):=m(L)\cap
 N_G^{sum}$ needs to be materialised into as many nodes
 as are mapped onto $u$ via $m$. The relationship of these
 materialised node to other nodes of the shape are inherited from the
 original shape graph. In addition, we have to decide whether to keep
 the summarised node out of which have made our materialisation, or
 to remove it. This represents the idea that summarised nodes
 can stand for any number of concrete nodes. Thus we get several
 materialisations of one shape graph, one for every set
 $I \subseteq \Gamma(m)$, $I$ being those
 now materialised nodes for which we keep the original summary node.

 \begin{definition}\label{def:materialisation}
  Let $G$ be a graph, $S=ls\br{G}=\abr{U,\mc P,\iota}$, $P=\abr{L,R}$
  be a production rule and $\mc M=\cbr{m\mid \sem{\varphi_P}_{\bar
  m}^S=\maybe}$. Let $\Gamma\br{m}:=m(L)\cap
  N_G^{sum}$. Then, for each $m\in\mc M$ and each
  $I\subseteq\Gamma\br{m}$ the \emph{materialisation of $P$ according
  to $\br{m,I}$} is defined as $mat_m^I\br{S}=\abr{U^I,\mc P,\iota'}$,
  with 
  \begin{equation*}
   U^I=U\setminus\br{m\br{N_L}\setminus I}\cup N_L
  \end{equation*}
  and for $p\in\binary$ and $q\in\unary\setminus\cbr{sm}$, letting
  $\widehat m=m\cup id_{U}$: 
  \begin{align*}
   \iota'(q)(u) &= \begin{cases} \true & \text{if } u \in
  N_L \wedge \br{u,q,u} \in E_L \\ \iota(q)\br{\widehat m\br{u}}
  & \text{else } \end{cases} \\ 
   \iota'(p)\br{u_1, u_2} &= \begin{cases} \true & \text{ if } u_1,
  u_2 \in N_L \wedge \br{u_1,p,u_2} \in E_L \\ \iota(p)\br{\widehat
  m\br{u_1}, \widehat m \br{u_2}} & \text{ else } \end{cases}\\ 
   \iota'(sm)(u) &= \begin{cases} \false & \text{ if } u\in
  N_L \\ \iota(sm)\br{\widehat m\br{u}} & \text{ else } \end{cases} 
  \end{align*}
  The collection of all such logical structures is then defined as
  the \emph{materialisation of $S$ with respect to $P$}: 
  \begin{align*}
   mat_P\br{S} &= \cbr{S\mid\exists m:\sem{\varphi_P}_m^S=\true} & \text{regular rule application}\\
               &\cup\cbr{mat_m^I\br{S}\mid \sem{\varphi_P}_m^S=\maybe,I\subseteq\Gamma\br{m}}
  & \text{materialisations} 
  \end{align*}
 \end{definition}

\noindent Note that the size of $mat_m^I$ can be exponential in the number of nodes
in the left hand side of the rule, but is finite. The following
theorem states that it is indeed sufficient to consider $mat_P\br{S}$
instead of $focus_P\br{S}$. 

\begin{theorem}
 Let $S$ be a $3$-valued logical structure and $P$ a production rule. Then
 \begin{align}
  mat_P\br{S} &\subseteq focus_P\br{S}\text{ and}\label{eq:mat_subset_focus}\\
  focus_P\br{S} &\lio mat_P\br{S}\label{eq:focus_embed_mat}
 \end{align}
\end{theorem}

\begin{figure}[htbp]
 \begin{center}
 	\includegraphics[width=\textwidth]{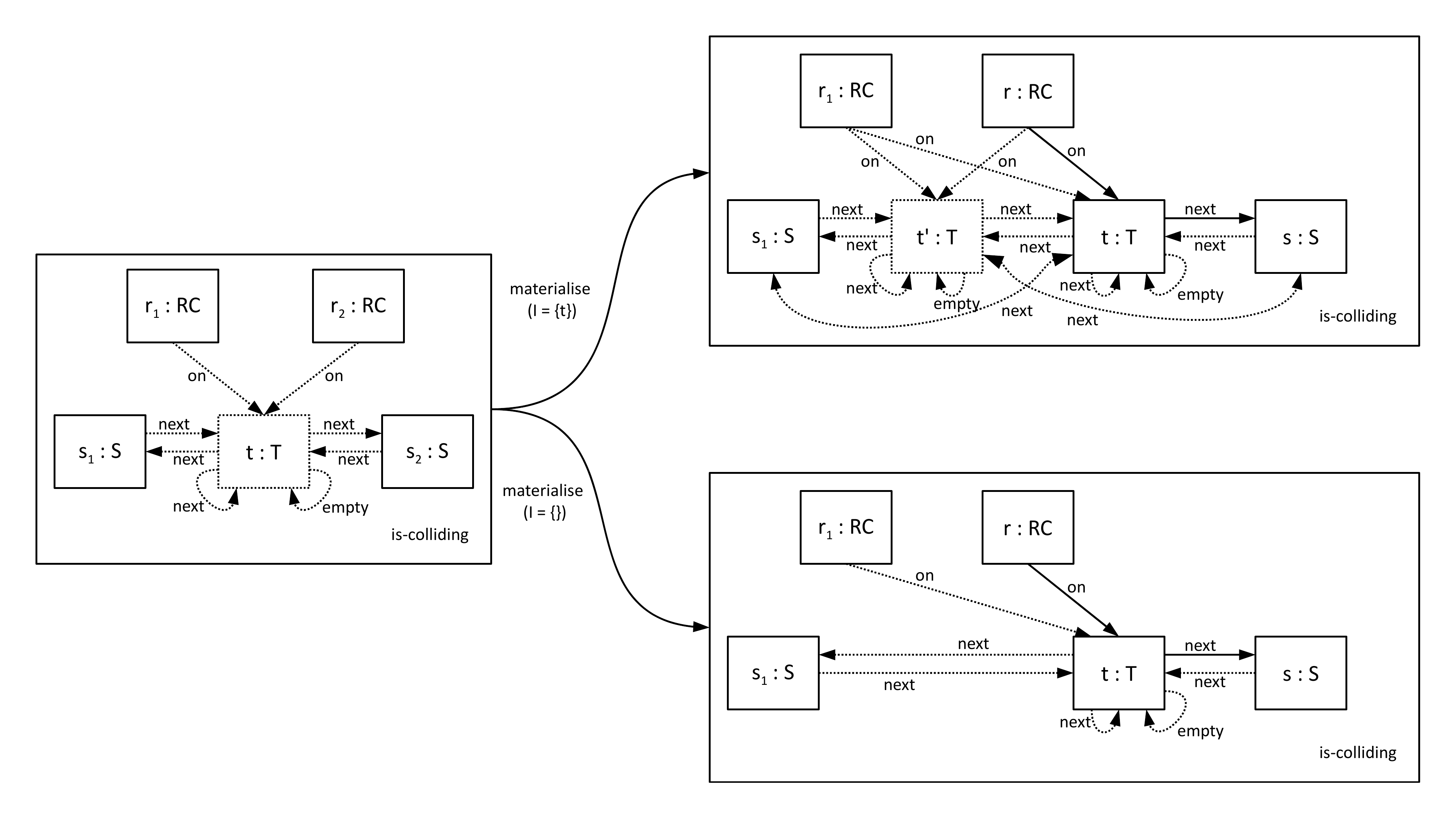}
 \end{center}
 \vspace{-1.5em}
 \caption{Materialisation of a shape graph wrt.\ $EnterStation$} 
 \label{fig:railcabmat}
 \vspace{-1.5em}
\end{figure}

\noindent Due to lack of space we have to omit all proofs. They can be
 found in \cite{SWW2010TR}. Fig.~\ref{fig:railcabmat} shows the
 result of applying materialisation on the starting shape graph using
 the $EnterStation$ production rule.  

The next step is {\em coercion}. After
 materialisation we apply the coerce operation defined
 in \cite{SagivRW02} on the resulting shape graphs. Doing so serves
 two purposes: On the one hand we can identify inconsistencies in the
 shape graph (e.g. an empty track with a railcab on it). On the other
 hand we can ``sharpen'' some predicate values of the shape graph in
 some cases. The latter can be found, for example, when looking at the
 materialised shape graphs of Fig.~\ref{fig:railcabmat}. There, the
 $empty$ predicate has the value $\maybe$ for node $t$. Yet, the
 railcab $r$ is definitely on it. Hence, we can sharpen the
 predicate value of $empty$ to $\false$. 
 
 The semantic knowledge needed to perform the coercion step comes from the so-called \emph{compatibility constraints}. Compatibility constraints may be either hand-written formulae (e.g. $\exists r: on(r, t) \Rightarrow \neg empty(t)$) or may be formulae that are derived from the so-called \emph{meaning formulae} of the instrumentation predicates (see below for an discussion of instrumentation predicates and its meaning formulae). We do not explain coercion in
 detail here, for this see \cite{SWW2010TR}. 

Finally, we can now apply the production rule. Since the left hand
side of the rule is - due to materialisation - explicitly present in
the shape graph, this follows the standard procedure. There is however
one speciality, related to the analysis, involved. To make the
analysis more precise, we introduce special {\em instrumentation
predicates} to our shape graphs. Consider again our forbidden
collision pattern of the last section. To see whether this is present,
we could evaluate the formula 

\begin{align*}
	\varphi_{forbidden\_collision} &:= on(r_1,t) \wedge on(r_2, t) \wedge T(t) \wedge RC(r_1) \wedge RC(r_2) \wedge \\
	&\phantom{= a} \neg (r_1 = r_2)
	\wedge \neg sm(r_1) \wedge \neg sm(r_2) \wedge \neg sm(t) 
\end{align*}
\noindent Unfortunately, for most shape graphs we find an assignment
$m$ such that this formula evaluates to $\maybe$ under $m$ since  
we have lost information about the precise position of railcabs on
tracks. This holds in particular also in our start shape graph. To regain this, we
introduce an extra instrumentation 
predicate for this property: $is\_colliding$. In our start shape
graph for the reachability analysis this predicate is $\false$ for all
nodes (see Fig.~\ref{fig:shape_graph_encoding} where the label $is\_colliding$ is not
connected to any node; we thus start the analysis with a shape in which no two
railcabs are on the same track). Every concretisation of a shape graph
with instrumentation 
predicates $p$ has to obey its so-called {\em meaning formula}
$\alpha_P$. For example the instrumentation predicate $is\_colliding$ has the following attached meaning formula:
\[ \alpha_{is\_colliding}(v) := T(v) \wedge \exists r_1, r_2: (r_1 \ne r_2 \wedge on(r_1, v) \wedge on(r_2, v)) \]
Now, for a concrete graph $G$ embedded in a shape $S$ with node $v$ mapped to $u$
via the embedding function, we have to check that the evaluation of
the meaning formula wrt.\ $v$ yields the same or a more precise value
(wrt.\ to the information order) than the instrumentation predicate
value in $u$. Instrumentation predicates are (obviously)
not part of our production  
rules. Therefore, we have to explicitly specify how these predicates 
change on rule application. For this purpose, we specify {\em update
formulae}. 

\begin{definition} 
A \emph{shape production rule} $P = (\lseq L,R \rseq, \gamma)$
consists of a graph production rule $\lseq L,R \rseq$ and function
$\gamma$ mapping from each instrumentation predicate
$p \in \mathcal{I}$ and each node $v \in N_R$ to a first-order
predicate-update formula $\varphi_{p,v}$ with free variables in $N_L$.  
\end{definition} 

\noindent The predicate-update formula $\varphi_{p,v}$ specifies how the value
of the instrumentation predicate $p$ should be calculated for each
$v \in N_R$ of the new shape graph with respect to the predicate
values of the old shape graph. For example, we could attach the
following update formulae to the production rule $EnterStation$: 
\begin{align*}
	\varphi_{is\_colliding, r} &= \false, \quad  
	\varphi_{is\_colliding, s} = \false \\
	\varphi_{is\_colliding, t} &= is\_colliding(t) \wedge {} \\
	&\phantom{= .} \exists r_2,r_3: ((r_2 \ne r) \wedge (r_3 \ne
	r) \wedge (r_3 \ne r_2) \wedge on(r_2, t) \wedge on(r_3, t)) 
\end{align*}

\noindent
Note that we make use of a free variable called $r$ in the formula
$\varphi_{is\_colliding, t}$. When the production rule is applied to a
shape graph $S$, this free variable gets assigned to the individual in
$S$ that represents the $r$ node of the left hand side of the
production rule. The following definition formalises the shape
production application. 

\begin{definition} 
	Let $P = (\lseq L,R \rseq, \gamma)$ be a shape production rule
	and $S = \lseq U,\iota \rseq$ be a shape graph. The rule
	$P$ \emph{can be applied} to $S$ if we find an injective
	function $m : N_L \rightarrow U$ (again called a matching)
	such that for all $(n,p,n') \in E_L$: $\iota(p)(m(n),m(n')) =
	1$ (or $\iota(p)(m(n)) = 1$ for $p \in \unary$). \\ 
	If $m$ is a matching, then the application of $P$ onto $S$
	with respect to the matching $m$ is the structure $S' = \lseq
	U', \iota'\rseq$ with $U' = 
	(U \setminus m(N^-)) \cup N^+$ and $\iota'$ defined as
	follows for $p \in \binary$, $o \in \unary \setminus \{sm\}$,
	$q \in \aux$, $\widehat{m} = m \cup \id_{N^+}$, and
	$u,u_1,u_2 \in U'$: 
	\[ \iota'(o)(u) = 
	\begin{cases}
		0 & \text{if } (u,o,u) \in m(E^-) \\
		1 & \text{if } (u,o,u) \in \widehat{m}(E^+) \\
		\iota(o)(u) & \text{else}
	\end{cases} \\
	\iota'(p)(u_1,u_2) = 
	\begin{cases}
		0 & \text{if } (u_1,p,u_2) \in m(E^-) \\
		1 & \text{if } (u_1,p,u_2) \in \widehat{m}(E^+) \\
		\iota(p)(u_1, u_2) & \text{else}
	\end{cases} \]
        \[\iota'(sm)(u) = \begin{cases} 0 & \text{if } u \in
	N^+ \\ \iota(sm)(u) & \text{else} \end{cases} \\ 
	\iota'(q)(u) = \begin{cases} \sem{\gamma(q,m^{-1}(u))}^{S}_{m}
	& \text{if } u \in m(N_L) \\ \sem{\gamma(q,u)}^{S}_{m}
	& \text{if } u \in N^+ \\ \iota(q)(u)
	& \text{else} \end{cases} \] 
	We write $S \trans{P, m} S'$ if $S'$ is the result of applying
	$P$ with matching $m$ to $S$.  
\end{definition} 

\begin{figure}[htbp]
 \begin{center}
 	\includegraphics[width=\textwidth]{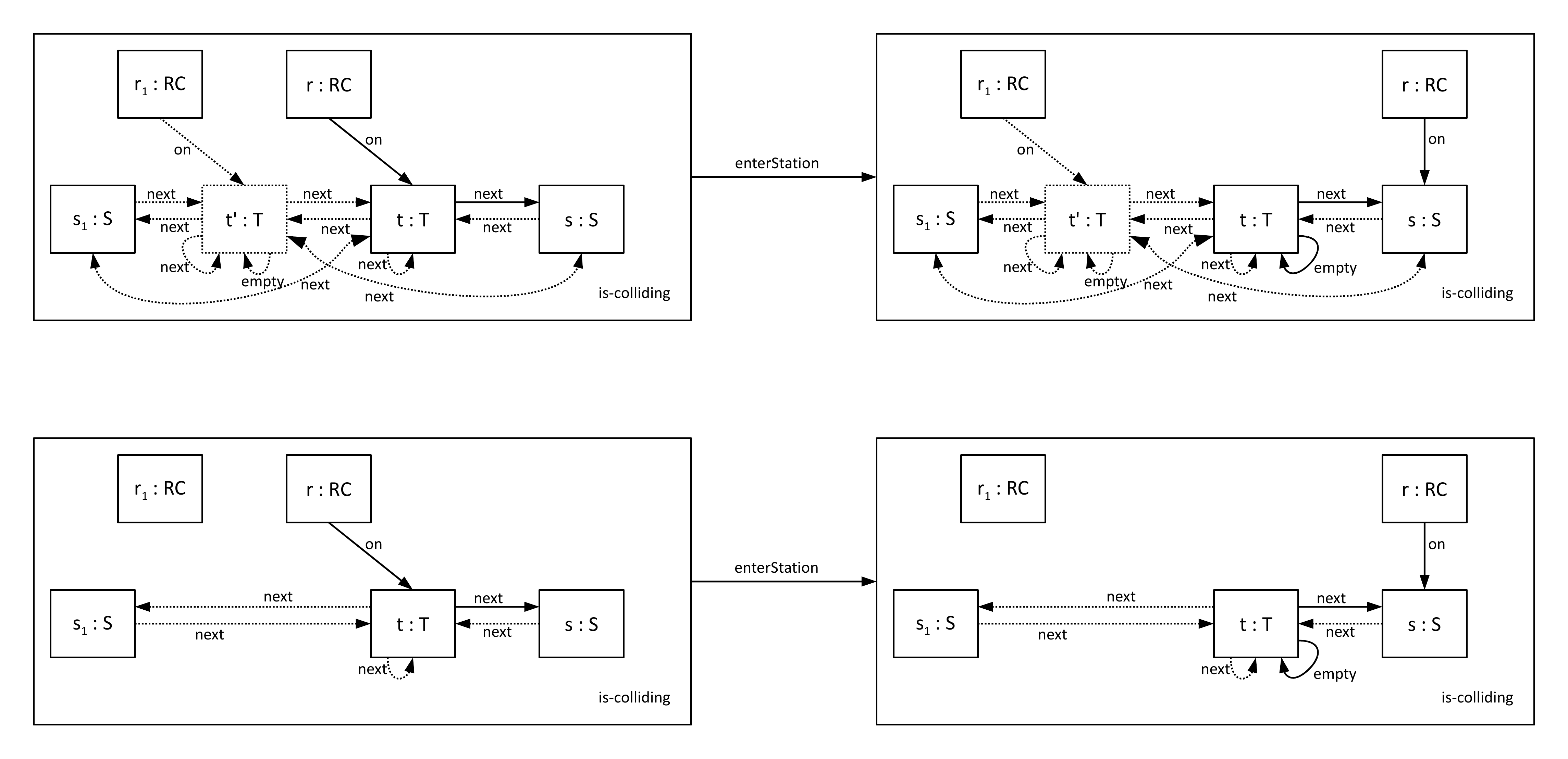}
 \end{center}
 \vspace{-1.5em}
 \caption{Applying rule $EnterStation$ on coerced shapes.} 
 \label{fig:railcabprodapp}
 \vspace{-2.5em}
\end{figure}

\noindent We also use the notation $\transo$ to include all steps of
	materialisation, coercion and rule application, i.e.\ we write
        $S \trans {} S'$ if $S$ can be materialised into $S_1$ wrt.\ a rule
        $P$, then coerced into $S_2$, $P$ applied giving
        $S_3$ and finally coerced into 
        $S'$. Figure~\ref{fig:railcabprodapp} shows the result of applying
$EnterStation$ on the coerced versions of the shapes of
Fig.~\ref{fig:railcabmat}.  

\section{Soundness of Technique}
 \label{sec:result}
 Using the methods of the previous section, we can now define the set
of reachable shapes of a shape graph transformation system $ST$, where
$ST$ consists of a start shape graph $S_0$ and a set of shape
production rules $(P_i, \gamma_i)_{i \in I}$:  

 \[ reach(ST) = \max \left( \{ S \mid S_0 \trans {}^* S \} \right )\]  

\noindent Here, $\max$ is defined for a set of shape graphs $XS$ as in
\cite{SagivRW02}: 
\[ \max(XS) := XS \setminus \{ X \mid \exists X' \in XS: X \sqsubseteq
X' \wedge X' \not \sqsubseteq X \} \] 

\noindent The set of reachable shape graphs can be inductively
constructed: we start with the initial shape graph and then
successively apply the production rules. For each newly produced shape
graph we check whether it can be embedded into or covers an already
existing shape graph. Shape graphs that are covered by others are
discarded.
%Whenever we reach a shape graph
%which is embedded in one of the already constructed shapes, we do not
%add it to our set of reachable graphs, otherwise we do. 
The following theorem states
that this algorithm is sound, i.e.\ we do not miss any of the reachable
graphs: 

\begin{theorem} Let $GT = (G_0,(P_i)_{i\in I})$ be a graph
  transformation system, $ST = (S_0, (P_i, \gamma_i))_{i \in I}))$ an
  associated shape transformation system with $G_0 \lio S_0$. Then 
  \[ reach(GT) \subseteq \{G \mid G \text{ 2-valued } \wedge G \lio S
  \wedge S \in reach(ST) \} \ .\] 
\end{theorem} 

\noindent  Note that due to lack of space we have left out
some extra conditions here referring to coercion and the compatibility
constraints used therein. The full theorem and the proof can be found
in \cite{SWW2010TR}.

At the end, we have to check for forbidden patterns in the shape
graphs. A shape graph $S$ {\em contains a  forbidden pattern}  $\lseq
F, F \rseq$ ($S \models F$) if (1) there is an assignment $m$ such
that $\sem{\varphi_F}_m^S \neq \false$, i.e.\ if the pattern is
(potentially) present in the shape, and materialisation and coercion
give us at least one valid concretisation, i.e.\ (2) $coerce(mat_P(S))
\neq \emptyset$.  If a forbidden pattern is not
contained in a shape graph, then it is also not contained in embedded
concrete graphs. 

\begin{theorem}
  Let $S$ be a shape graph, $\lseq F,F \rseq$ a forbidden pattern, $G$
  a graph such that $G\lio S$. Then $S \not \models F \implies G \not
  \models F$. 
\end{theorem}

\noindent In summary this shows soundness of our technique: all
reachable graphs are embedded in reachable shape graphs, and if we are
able to show absence of forbidden patterns in the shapes this also
holds for the concrete graphs. Note that due to the overapproximation
the reverse is in general not true: we might find forbidden patterns
in the shapes although none of the concrete graphs contain
them. Instrumentation predicates are used to reduce such
situations. 

Finally, a note on termination. If the algorithm is carried out as
proposed above, it might not terminate although we only consider
maximal shapes. This could occur if the production rules generate
shapes which are all incomparable in the embedding order. To avoid
this, one can introduce another abstraction step in the algorithm:
Nodes which agree on all unary predicate 
valuations are collapsed into one. As we can only have finitely many
combinations of predicate valuations this gives us finitely many
different shape graphs.

\section{Implementation}
 We implemented the verification algorithm  in Java, making use of the
source code of the 
shape analysis tool TVLA \cite{BogudlovLRS07}. Thus we were
able to take advantage of the already optimised code for logical
structures provided by TVLA. Basically, our implementation loads a
starting shape graph, a set of 
shape production rules, and a set of forbidden patterns, represented
as text files each. Additionally, one needs to supply a text file
listing the set of core and instrumentation predicates, the latter
with their meaning formulae. The implementation then
successively constructs the set of reachable shape graphs, each
represented as logical structure, and checks whether a newly found
shape graph contains one of the forbidden patterns. If the shape
graph does contain a forbidden pattern, a counter example is generated
that describes how the shape graph was constructed as sequence of
production applications. Otherwise, the shape graph is added to the
set of reachable shape graphs and the maximum operation is applied. If
no new shape graphs can be found anymore and none of the reachable
shape graphs contains a forbidden pattern, the implementation asserts
that the given STS is safe. 

We tested our implementation using the running example on a 3GHZ Intel
Core2Duo Windows System with 3GB main memory. Our implementation needs
about 250ms to verify that the running example STS is safe, i.e.\ no collision
happens.  While doing so, it
temporarily constructs 108 intermediate logical structures and finds
17 logical structures in the maximised set of reachable shape graphs. 

This and further case studies show that the most expensive operation in terms of
runtime is the $max$ operation. We implemented it by
checking for each newly found shape graph whether it can be embedded
in a shape graph in the (current) set of reachable shape graph or vice
versa. Thus, for each newly found shape graph we need $2n$ embedding
checks, if $n$ denotes the number of shape graphs in the current set
of reachable shape graphs. Furthermore, for arbitrary shape graphs
checking for embedding is NP-complete
(\cite{Arnold06combiningshape}). Hence it is not surprising that the
$max$ operation was observed to be very costly.
 
\section{Conclusion and Related Work}

\label{sec:conclusion}
In this paper, we have introduced a shape analysis approach for generating a finite over-approximation of the reach set of a graph transformation system with infinite state space. In contrast to some of the other work done in this area, e.g. \cite{Rensink04}, we derive from our strict adherence to the formalism presented in \cite{SagivRW02} a very straightforward avenue for implementation, which we have demonstrated using the $3$-valued logic engine TVLA. In order to emphasize the qualities of our approach, we will now discuss how it relates to other work in this area.

In \cite{BauerBKR08}, a method for automatic abstraction of graphs is introduced. Intuitively, nodes are identified if their neighbourhood of radius $k\in\mathbb N$ is the same. While this automatic abstraction greatly reduces the need for human intervention in the verification process, it also reduces the flexibility of the approach. Only a certain class of systems can be handled well by neighbourhood abstraction, while our approach can be tuned to fit the needs of very different systems on a per-system basis. Furthermore, the method from \cite{BauerBKR08} cannot use information from spurious counterexamples, since the abstraction leaves them with only one degree of freedom, the radius $k$. In contrast, using additional instrumentation predicates, our approach can utilise the full amount of information from spurious counterexamples.

Another approach to verifying infinite-state systems is the one by Saksena, Wibling and Jonsson \cite{SWJ2008}. It is based on backwards application of rules. By applying inverted rules to the forbidden patterns it is possible to determine whether a starting graph can lead to a failure state. The backwards application paradigm imposes some restrictions on this approach, for example forbidding the deletion of nodes and requiring a single starting pattern. Our approach does not suffer such restrictions. Furthermore, since the approach does not include an explicit abstraction and thus no information about the rest of the graph is available when applying a rule to a pattern, it would be very difficult to include concepts such as parameterised rules or parallel rule application. Since our approach uses explicit abstraction through shapes, it can encode information about the entire graph and is thus much more suited to support such extensions.

Lastly, Baldan, Corradini and König \cite{BaldanCK08} have written a series of papers in which they develop a unique approach to the verification of infinite state GTSs. They relate GTSs to Petri nets and construct a combined formalism, called a petri graph, on which they show certain properties via a technique called unfolding. This approach achieves many of the goals we strive for. However, a single concrete start graph is required for an analysis, which would be a major restriction in systems where there are many possible initial states, or even an unknown initial state.

The above discussion of related work is by no means exhaustive, but it suffices to show that, while each of these approaches has currently some advantages over our approach, no single approach outperforms ours in every single way. The results described in this paper lay the foundations for a new approach to the verification of infinite-state GTSs, which we strongly believe to be better suited to overcome the many problems facing any theory in this area, than the currently available approaches. As such, there are a number of limitations to our approach which we intend to tackle in the future.  We plan to look at parallel rule application, negative application conditions \cite{HabelHT96} and especially rules with quantifiers \cite{rensink2006nested} which allow to specify changes on arbitrary numbers of nodes of some particular type within one rule.

\bibliographystyle{alphaabb}
\bibliography{references}

\end{document}